\documentstyle[12pt]{article}
\def\be{\begin{equation}}
\def\ee{\end{equation}}
\def\ba{\begin{array}}
\def\ea{\end{array}}
\def\bea{\begin{eqnarray}}
\def\eea{\end{eqnarray}}
\def\br{\begin{eqnarray}}
\def\er{\end{eqnarray}}

\begin{document}
\begin{center}
{\LARGE \bf Decay studies of $^{288-287}115$ alpha-decay chains}\\
\bigskip
{\bf Sushil Kumar$^{1,*}$, Shagun Thakur$^{2}$, Rajesh Kumar$^{2}$}\\

\medskip

$^{1}$ Department of Physics, Chitkara University, Solan-174103, H. P., India\\
$^{2}$ Department Physics, National Institute of Technology, Hamirpur, H. P.,India\\

\medskip
\end{center}
\vspace*{0.3cm}
\begin{center}
{\bf Abstract}
\end{center}
\medskip
\baselineskip 18pt
The $\alpha$-decay chains of $^{288-287}115$ are studied
along with the possible cluster decay modes by using the preformed cluster
model (PCM). The calculated $\alpha$-decay half-lives are compared with experimental
data and other model calculations. The calculated Q-values, penetration probabilities
and preformation  probabilities factors for $\alpha$-decay suggest that $^{283}_{170}113$,
$^{287}_{172}115$ and $^{272}_{165}107$ parent nuclei are  more stable against
the $\alpha$-decay. These alpha decay chains are further  explored for the
possibilities of cluster decay. Decay half lives of different cluster from
different nuclei of the decay chains point to the extra stability near or
at the deformed shells $Z=108$, $N=162$ and $Z=100$, $N=152$. The decay half-lives
for $^{14}C$ and $^{48}Ca$ clusters are lower than the current experimental
limit ($\approx$ $10^{28}$sec).

\vspace*{2.0cm}
* sushilk17@gmail.com
\vfill\eject
\section {\large \bf Introduction}
The synthesis of superheavy elements and their decay studies is a long
term goal of nuclear structure physics. These days advancement in the
radioactive nuclear beam facilities has opened the door to reach the center
of island of superheavy elements. Understanding nuclear stability/instability
in the superheavy mass region is also a long standing question. Firstly in
$1969$ Flerov \cite{ref1} suggested the use of a highly neutron-rich beam
of $^{48}Ca$ for the formation of superheavy elements with neutron rich targets
such as $^{244}Pu$, $^{248}Cm$ and $^{252}Cf$.\\
Many superheavy elements have been synthesised at GSI, Dubna, RIKEN and LBNL.
The progress of the synthesis of odd $Z$ nuclei is rather slow as compared
to even $Z$ nuclei. The nuclei in superheavy mass region are investigated
through the measurement of energies of alpha particles belonging to the
characteristic chains of such superheavy elements. The most important decay
parameters for these decay chains are Q-value of the decay and the half-life
time. Extensive attempts have been made to calculate the alpha decay half
lives theoretically. Alpha decay half-life are very sensitive to the Q-values
and the Q-values depend on the mass of parent and daughter nuclei. Different
approaches have been adopted to find the nuclear mass in the superheavy region.
In recent years, a  number of papers in which alpha decay has been studied
by determining the Q-value from different mass tables and then putting it
into different formalisms to calculate alpha decay half lives have appeared
\cite{ rref1, rref2, rref3, rref4, rref5}. In \cite{ rref6, rref7}Attempts
to find the properties of the nuclei that belong to the alpha decay chain
of the superheavy elements have also been made.\\
Recently two alpha decay chains $^{288}115$ and $^{287}115$ have been
observed \cite{ ref2} in the 3n- and 4n- evaporation channels with cross
sections of about 3pb and 1pb respectively, using the $^{48}Ca$ beam with
$^{243}$Am target. These two decay chains have been studied \cite{ zhongzhou03,
geng03, sankha04, basu04, sharma05, royer08} for the ground state properties
using various theoretical models. The purpose of this work is to study the
alpha decay and cluster decay half-lives of superheavy nuclei and the shell
closure effects if any, within the Preformed Cluster Model (PCM). In the
first part of the calculations alpha decay half-lives are calculated for
both the alpha decay chains. Second part of the calculation covers comparative
study of alpha decay half-lives for both $^{288-287}115$ decay chains with
other theoretical model calculations and also calculation within the framework
of PCM model by taking different Q-values from experimental data i.e.PCM
model calculation, Myers-Swiatecki and from Muntian nuclear data table. These
calculated half-lives have been compared with the experimental results. Heavy
nuclei of trans lead region are well known alpha emitters and these nuclei
also undergo exotic cluster decay that was predicted by Poenaru et.al. in
$1980$ \cite{rref10}. After the first experimental observation of cluster
decay by Rose and Jones \cite{basuclus2}, different approaches have been
followed to understand cluster decay \cite{ rref9, basuclus5}
and it is very well established mode of decay. Finally the theoretical calculations
for the emission of such exotic cluster heavier than alpha particle from
the nuclei belonging to the alpha decay chain of superheavy elements have
been made.\\

The Preformed Cluster Model (PCM) is described briefly in Section $2$
and the results of calculation are presented in Section $3$. Discussion and
summary of results is given in Section $4$.\\

\section {The preformed cluster model}
The Preformed Cluster Model (PCM) \cite{gupta88, malik89, kumar97}
uses the dynamical collective coordinates of mass and charge
asymmetries
$$\eta={{(A_1-A_2)}/{(A_1+A_2)}}$$
and
$$\eta _Z={{(Z_1-Z_2)}/{(Z_1+Z_2)}},$$
first introduced in the Quantum Mechanical Fragmentation Theory
\cite{gupta99a,maruhn74,gupta75,gupta77a,gupta77b,gupta77c}. These
are in addition to the usual coordinates of relative separation R
and deformations $\beta_i$ ($i=1,2$). Then, in the standard
approximation of decoupled R- and $\eta $-motions
\cite{gupta88,malik89,gupta91}, The decay constant $\lambda $  (the
decay half-life $T_{1/2}$ ) in PCM is defined as \be \lambda ={{{ln
2}\over {T_{1/2}}}}=P_0\nu _0 P. \label{eq:1} \ee Here $P_0$ is the
cluster (and daughter) preformation probability and P the barrier
penetrability which refer, respectively, to the $\eta$ and R
motions. The $\nu _0$ is the barrier assault frequency. The $P_0$
are the solutions of the stationary Schr\"odinger equation in
$\eta$, \be \{ -{{\hbar^2}\over {2\sqrt B_{\eta \eta}}}{\partial
\over {\partial \eta}}{1\over {\sqrt B_{\eta \eta}}}{\partial\over
{\partial \eta }}+V_R(\eta )\} \psi ^{({\nu})}(\eta ) = E^{({\nu})}
\psi ^{({\nu})}(\eta ), \label{eq:2} \ee which on proper
normalization are given as \be P_0={\sqrt {B_{\eta \eta}}}\mid \psi
^{({0})}(\eta (A_i))\mid ^2\left ({2/A}\right ), \label{eq:3} \ee
with i=1 or 2 and $\nu$=0,1,2,3.... Eq. (\ref{eq:2}) is solved at a
fixed $R=R_a=C_t(=C_1+C_2)$. The $C_i$ are S\"ussmann central radii
$C_i=R_i-({1/R_i})$, with the radii
$R_i=1.28A_i^{1/3}-0.76+0.8A_i^{-1/3} fm$.

The fragmentation potential $V_R(\eta )$ in (\ref{eq:2}) is
calculated simply as the sum of the Coulomb interaction, the nuclear
proximity potential \cite{blocki77} and the ground state binding
energies of two nuclei, \be V(R_a, \eta) =- \sum_{i=1}^{2} B(A_{i},
Z_{i})+ \frac{Z_{1} Z_{2} e^{2}}{R_a} + V_{P}, \label{eq:4} \ee with
B's taken from the 2003 experimental compilation of Audi et al.
\cite{audi03} and from the 1995 calculations of M\"oller et al.
\cite{moller95} whenever not available in \cite{audi03}. Thus, full
shell effects are contained in our calculations that come from the
experimental
and/or calculated \cite{moller95} binding energies. \\

The charges Z$_1$ and Z$_2$ in (\ref{eq:4}) are fixed by minimizing
the potential in $\eta_Z$ coordinate. The Coulomb and proximity
potentials in (\ref{eq:4}) are for spherical nuclei. The mass
parameters $B_{\eta \eta}(\eta )$, representing the kinetic energy
part in (\ref{eq:2}), are the classical hydrodynamical
masses \cite{kroeger80}. \\

The WKB tunnelling probability, calculated is  $P=P_i P_b$ with \be
P_i=exp[-{2\over \hbar}{{\int }_{R_a}^{R_i}\{ 2\mu [V(R)-V(R_i)]\}
^{1/2} dR}] \label{eq:5} \ee \be P_b=exp[-{2\over \hbar}{{\int
}_{R_i}^{R_b}\{ 2\mu [V(R)-Q]\} ^{1/2} dR}]. \label{eq:6} \ee These
integrals are solved analytically \cite{malik89} for $R_b$, the
second turning point, defined by $V(R_b)=Q$-value for the
ground-state decay. The assault frequency $\nu _0$ in (\ref{eq:1})
is given simply as \be \nu _0=(2E_2/\mu )^{1/2}/R_0, \label{eq:7}
\ee with $E_2=(A_1/A) Q$, the kinetic energy of the lighter
fragment, for the
$Q$-value shared between the two products as inverse of their masses.\\

\section {\large \bf Calculations and Results}
In Fig.1 decay characteristics of both alpha decay chains $^{288}115$
and $^{287}115$ are shown. The maximum half-life for the  alpha decay
chain of $^{287}115$ is found at $^{283}_{170}113$, $^{287}_{172}115$ parent
nuclei indicating that these are more stable against the alpha decay.
This stability can be attributed to either the magicity
of protons at $Z=114$ or of neutrons at $N=172$ or to both. The alpha decay
half-life for $Z=107$, $N=165$ is very high in alpha decay chain of $^{288}115$.
Smaller Q-value should mean a relative decrease in the penetrability P, shown
in Fig. 1. The preformation factor $P_{0}$ is smaller for $^{283}_{170}113$,
$^{287}_{172}115$ and $^{272}_{165}107$ parent nuclei and in present calculations
it is smaller as compared to those for the actinides \cite{sushil09}. There
is discontinuity for the half-lives between $^{279}111$ and $^{283}113$.
Such discrepancies in the decay of superheavy nuclei have been reported earlier
also \cite{rref11}. One of the reasons may the assumption that we are considering
alpha decay from the ground state of the parent to the ground state of daughter,
whereas the possibility of alpha decay from excited state also exists \cite{rref12}
in addition to this there is very large uncertainty in the experimental data
for these two nuclei. Further experimental data would be useful in understanding
this behaviour. In Fig. 2 the $PCM$ based calculations for $^{288}115$ and
$^{287}115$ alpha decay half-lives are compared with the experimental data.
The calculations based on other theoretical models are shown in the figure
and these are also given in Table 1. The calculated results in $PCM$ in 
general follow the trend of experimental data as is evident from the figure.\\
We took Q-values from experimental data($Q_{exp}$)\cite{ref2} and from different
models i.e. Myers-Swiatecki($Q_{MS}$)as well as Muntian et al. nuclear data
table($Q_{MU}$) from Ref.\cite{rref2}, whereas $PCM$ ($Q_{PCM}$) has been
calculated using Audi et al.\cite{audi03} and Moller et al. nuclear data
table\cite{moller95}. Using these Q-values we calculated $\alpha$-decay half-lives
for $^{288}115$ decay chain within the framework of PCM model. The results of these calculations are shown in Fig.3. These calculated half-lives are compared with the experimental results. In
this analysis it is found that calculated $\alpha$-decay half-lives using
Muntian's $Q_{MU}$-value in PCM improves the agreement with the experimental
results for this decay chain. Similar comparison for $^{287}115$-decay
chain is presented in Fig.4.\\
The second part of the calculations is done to look for any possibility of
cluster decays from these two $^{288-287}115$ alpha decay chains. In Fig.
5, the calculated cluster decay half-lives of various possible clusters are
shown with the Q-values for $^{288}115$. The choice of the clusters is based
on the minima in the fragmentation potentials $V(\eta)$\cite{sushil03}. The
Q-value increases as the size of the cluster increases, its increase with
the mass of parent nucleus is smooth and linear.
The study of half lives of different cluster decay modes tells about the
shell effects. Higher value of half life indicate the presence of shell
stabilised parent nucleus, whereas comparatively low value of half life tells
the same about the daughter and cluster  nuclei. A careful analysis of different
cluster emissions from the nuclei of alpha decay chain of $^{288}115$ shows
that the shell stabilising effect is seen in the decay of $^{10}Be$ from
$^{272}107$ nucleus since decay half life for this cluster peaks at this
parent nucleus. Also in $^{10}Be$, $^{14}C$, and $^{26}Ne$ decay of $^{280}111$
nucleus a minimum is seen therefore these decays lead to deformed daughters
with (Z, N) values as (107, 163), (105, 161) and (101, 153) respectively.
It points to extra stability near the deformed shells ($Z=108$, $N=162$)and
($Z=100$, $N=152$) as investigated in \cite{lazarev94, greenless08}. 
The cluster preformation ($P_{0}$) and penetration (P) probabilities are
shown in Fig. 6. Since assault frequency ($\nu_{0}$)remains almost constant,
the decay half-life is a combined effect of preformation ($P_{0})$ and penetration
(P) probabilities. From Fig. 6 it is clear that $^{10}Be$ has higher $P_{0}$
as compared to clusters $^{14}C$, $^{20}O$, $^{23}F$, $^{26}Ne$, $^{28}Mg$,
$^{34}Si$ and  $^{48}Ca$ but its penetration probability is smaller as compared
to different clusters yet it has larger half-life than the others. Similar
calculations made for the second alpha decay chain $^{287}115$ are shown
in Fig. 7 and Fig. 8. In Fig. 7, similar to Fig. 5 prominent shell effects
are seen in some of the cluster decay modes. High value of  half life for
the decay of $^{10}Be$ from  $^{271}107$ nucleus may be due to deformed magicity
of $Z=107$ and $N=164$. Also the presence of minimum in the half life of
$^{14}C$, and $^{24}Ne$ decay from $^{279}111$ nucleus points to the extra
stability of the daughter with $Z=105$, $N=160$ and $Z=101$, $N=152$. Interestingly
it is found that decay half-life of $^{48}Ca$ cluster from all parent nuclei
of this decay chain lies within the experimental limits $\sim$$10^{28}$s\cite{gupta94}.\\

\section{Summary}
In this paper PCM based calculation have been used study alpha
and cluster decay of two decay chains of $^{287}115$ and $^{288}115$ nuclei.
A comparative study of both the $\alpha$-decay chains has been carried first.
Alpha decay half-lives are compared with the experimental data and the other
available theoretical calculations. The PCM calculation results in general
follow the experimental data.\\
Not only PCM calculation results are compared with the experimental
data and other theoretical models but also the role of Q-value
in half-life is studied. Then we calculated the half-life using PCM
model by taking the Q-values from experimental data ($Q_{exp}$), PCM
model($Q_{PCM}$), Myers-Swiatecki ($Q_{MS}$) and from Muntian ($Q_{MU}$)nuclear
data table. It is found that the use of ($Q_{MU}$) provides better agreement with the experimental results. Also, a small change in Q-value produce a large variation in half-life, which indicates the sensitivity of half-lives towards the Q-values.\\
Finally, cluster decay calculations made for both $\alpha$-decay chains $^{288-287}115$
with a view to see if there is any branching of the $\alpha$-decay to another
light nucleus due to the spherical and/or deformed magicity of the corresponding
heavy daughter nucleus. After a careful analysis of the two alpha decay chains
$^{288-287}115$, which offers some possibilities of cluster emission such
as $^{10}Be$, $^{14}C$, $^{26}Ne$ from $^{279-280}111$. The decay half-lives
of $^{14}C$ cluster decay from $^{284}113$ and $^{280}111$,  $^{48}Ca$ from
$^{288}115$ and $^{284}113$ lies within the range of experimental limits.\\

\newpage
\par\noindent
{\bf Figure Captions}\\

\par\noindent

Fig. 1 {The $\alpha$-decay half-lives, Q-values, preformation and the penetration
probabilities calculated on the basis of PCM plotted as a function of the parent
nucleus charge for the $\alpha$-decay chains of $^{288-287}115$.}\\

\par\noindent
Fig. 2 {The $\alpha$-decay half-lives calculated on the basis of PCM
and comparison with experimental data\cite{ref2} and those calculated
on the basis of the Delta2\cite{geng03}, NL3 parameter calculation\cite{sankha04},
DDM3Y effective interaction\cite{basu04}, GLDM and VSS from\cite{royer08}
plotted as a function of the parent nucleus mass  for the $\alpha$-decay chains of $^{288-287}115$.}\\

\par\noindent
Fig. 3 {The $\alpha$-decay half-lives and the Q-values are plotted as a
function of the parent nucleus mass for the $\alpha$-decay chain of $^{288}115$.
The $\alpha$-decay half-lives are calculated using experimental Q-values
($Q_{exp}$)\cite{ref2} and theoretical Q-values from the PCM model
calculation($Q_{PCM}$), Myers-Swiatecki($Q_{MS}$) and from Muntian et al.
nuclear data table($Q_{MU}$). The ($Q_{MS}$,$Q_{MU}$) values are taken from
Ref.\cite{samanta07} while($Q_{PCM}$) is calulated using the Audi-Wapstra mass
table\cite{audi03} and M\"oller et al. nuclear data table\cite{moller95}.}\\

\par\noindent
Fig. 4 {The same as for Fig. 3 but for the $^{287}115$ $\alpha$-decay chain.}\\

\par\noindent
Fig. 5 {The calculated half-lives and the Q-values for different cluster
decays on the basis of PCM model plotted for the parent nuclei belonging
to $\alpha$-decay chain of $^{288}115$.}\\

\par\noindent
Fig. 6 {The same as for Fig. 5 but for the preformation probability $P_{0}$ and the penetration probability P.}\\

\par\noindent
Fig. 7 {The same as for Fig. 5 but for the $^{287}115$ $\alpha$-decay chain.}\\

\par\noindent
Fig. 8 {The same as for Fig. 6 but for $^{287}115$ $\alpha$- decay chain}\\

\newpage

\begin{table}
\caption{Comparision between experimental and calculated PCM
$\alpha$-decay half-lives with other theoretical calculations.
Calculated results are taken from Delta2 \cite{geng03}, NL3
parameter calculation \cite{sankha04}, DDM3Y effective
interaction\cite{basu04}, GLDM and VSS from \cite{royer08}.
$Q_{PCM}$ is calculated using binding energies from
Audi-Wapstra\cite{audi03} and M\"oller et al. nuclear data
table\cite{moller95}.}
\begin{tabular}{|c|c c|c c c c c c c|}\hline
\multicolumn{1}{|c|} {} &\multicolumn{2}{|c|}
{Q(Mev)}&\multicolumn{7}{|c|} {$log_{10}T_{1/2}$ (sec)} \\\hline
$^{A}Z$&Exp.&PCM&Exp.& PCM &DDM3Y & GLDM &VSS&Delta2&NL3 \\ \hline
$^{288}115$&10.61 &10.99&-1.06&0.27&-0.39&-1.02&-0.0013&0.84&\\
$^{284}113$&10.15&10.25&-0.32&1.56 &0.187&-0.37&0.62&-0.95& \\
$^{280}111 $&9.87 &9.98&0.56&1.82 &0.28&-0.16&0.76&-0.93&\\
$^{276}109$&9.85&9.80&-0.14&1.93  &-0.35&-0.72&0.16&1.40&\\
$^{272}107$&9.15 &9.30&0.99&2.86  &0.99&&1.52&3.29&\\
$^{287}115$&10.74&11.30&-1.45&2.35 &-1.31&-1.33&-0.69&-1.09 &-1.12\\
$^{283}113$&10.26&10.60&-1.00&2.61&-0.69&-0.65&-0.03&-1.89&1.06\\
$^{279}111$&10.52&10.45&-0.77&-0.41&-2.00&-1.92&-1.35&-0.85&-0.89\\
$^{275}109$&10.48&10.12&-2.01&-0.06&-2.57&-2.39&-0.86&0.76&0.08\\
$^{271}107$&-&9.50&-&1.22&0.65&-0.30&1.43&2.78&2.31\\\hline

\end{tabular}
\end{table}


\newpage

\begin{thebibliography}{99}
\bibitem{ref1}
G. N. Flerov,  Atom. Ener. 26, 138 (1969).
\bibitem{rref1}
P. R. Chowdhury, and C. Samanta, At. Dat. Nucl. Dat. Tables, {\bf 94},781
(2008).
\bibitem{rref2}
C. Samanta {\it et al.}, Nucl. Phys. {\bf A 789},142 (2007).
\bibitem{rref3}
D. N. Poenaru {\it et al.}, Phy. Rev. {\bf C74}, 014312 (2006).
\bibitem{rref4}
P. R. Chowdhury {\it et al.}, Phy. Rev. {\bf C73}, 014612 (2006).
\bibitem{rref5}
D. N. Poenaru {\it et al.} J. Phys. G: Nucl. Part. Phys. {\bf 32}, 1223 (2006).
\bibitem{rref6}
H. Zhang {\it et al.}, Phy. Rev. {\bf C71}, 054312 (2005).
\bibitem{rref7}
O. Parkhomenko, A. Sobiczewski, Acta Phys. Pol. {\bf B36}, 1363 (2005).
\bibitem{ref2}
Yu. Ts. Oganessian {\it et al.}, Phys. Rev. {\bf C 69}, 021601(R)
(2004).
\bibitem{zhongzhou03}
Zhongzhou Ren {\it et al.}, Phys. Rev. {\bf C67}, 064312 (2003).
\bibitem{geng03}
L. S. Geng {\it et al.}, Phys. Rev. {\bf C 68}, 061303(R)
(2003).
\bibitem{sankha04}
Sankha Das and G. Gangopadhyay, J. Phys. G: Nucl. Part. Phys. {\bf
30}, 957 (2004).
\bibitem{basu04}
D. N. Basu, J. Phys. G: Nucl. Part. Phys. {\bf 30}, B35 (2004).
\bibitem{sharma05}
M. M. Sharma {\it et al.}, Phys. rev. {\bf C
71}, 054310 (2005)
\bibitem{royer08}
G. Royer and H. F. Zhang, Phys. Rev. {\bf C77}, 037602 (2008).
\bibitem{rref10}
A. Sanduluscu, D. N. Poenaru and W. Greiner, Sov. J. Part. Nucl. {\bf 11},
528 (1980).
\bibitem{basuclus2}
H. J. Rose and G. A. Jones, Nature (London) {\bf 307}, 245 (1984).
\bibitem{rref9}
D. N. Poenaru, {\it et al.} At. Dat. Nucl. Dat. Tables, {\bf 48},231 (1991).
\bibitem{basuclus5}
B. Buck, {\it et al.}, At. Dat. Nucl. Dat. Tables, {\bf 54},53 (1993).
\bibitem{gupta88}
R.K. Gupta, in {\it Proceedings of the 5th International Conference
on Nuclear Reaction Mechanisms}, Varenna, Italy (1988), Editor: E.
Gadioli, Ricerca Scientifica ed Educazione Permanente, Milano, p.
416 (1988).
\bibitem{malik89}
S.S. Malik and R.K. Gupta, Phys. Rev. {\bf C39},1992 (1989).
\bibitem{kumar97}
S. Kumar and R.K. Gupta, Phys. Rev. {\bf C55},218 (1997).
\bibitem{gupta99a}
R.K. Gupta and W. Greiner, in {\it Heavy Elements and Related New
Phenomena}, Editors: W. Greiner and R.K. Gupta, World Sc., Vol. I,
p. 397; 536 (1999).
\bibitem{maruhn74}
J. Maruhn and W. Greiner, Phys. Rev. Lett. {\bf 32},548 (1974).
\bibitem{gupta75}
R.K. Gupta, W. Scheid and W. Greiner, Phys. Rev. Lett. {\bf 35},353
(1975).
\bibitem{gupta77a}
R.K. Gupta, C. P\^ arvulescu, A. S\u andulescu and W. Greiner, Z.
Physik {\bf A283},217 (1977).
\bibitem{gupta77b}
R.K. Gupta, A. S\u andulescu and W. Greiner, Z. Naturforsch. {\bf
32a},704 (1977).
\bibitem{gupta77c}
R.K. Gupta, Sovt. J. Part. Nucl. {\bf 8},289 (1977); Nucl. Phys. and
Solid St. Phys. Symp. (India) {\bf 21A},171 (1978).
\bibitem {gupta91}
R.K. Gupta, W. Scheid, and W. Greiner, J. Phys. G {\bf 17},1731
(1991).
\bibitem{blocki77}
J. Blocki, J. Randrup, W.J. Swiatecki, and C.F. Tsang, Ann. Phys.
(NY) {\bf 105},427 (1977).
\bibitem {audi03}
G. Audi, A.H.Wapstra and C.Thibault, Nuclear Physics {\bf A729},337
(2003).
\bibitem{moller95}
P. M\"oller, J.R. Nix, W.D. Myers and W.J. Swiatecki, At. Data Nucl.
Data Tables, {\bf 59},185 (1995).
\bibitem{kroeger80}
H. Kr\"oger and W. Scheid, J. Phys. G {\bf 6},L85 (1980).
\bibitem{sushil09}
Sushil Kumar {\it et al.}, J. Phys. G:Nucl. Part. Phys. {\bf 36},015110
(2009).
\bibitem{rref11}
C.Smanta, Porg. In Part. and Nucl. Phys. {\bf 62}, 344 (2009).
\bibitem{rref12}
D. S. Delion, {\it et al.}, Phy. Rev. {\bf C76}, 044301 (2007).
\bibitem{sushil03}
Sushil Kumar {\it et al.}, J. Phys. G:Nucl. Part. Phys. {\bf 29}, 625
(2003).
\bibitem{lazarev94}
Yu. A. Lazarev {\it et al.}, Phy. Rev. Lett. {\bf 73},624 (1994).
\bibitem{greenless08}
P. T. Greenless {\it et al.}, Phy. Rev. {\bf C78}, 021303(R) (2008).
\bibitem{gupta94}
R. K. Gupta and W. Greiner, Int. J. Mod. Phys. E (Suppl), {\bf3},335
(1994).


\end{thebibliography}
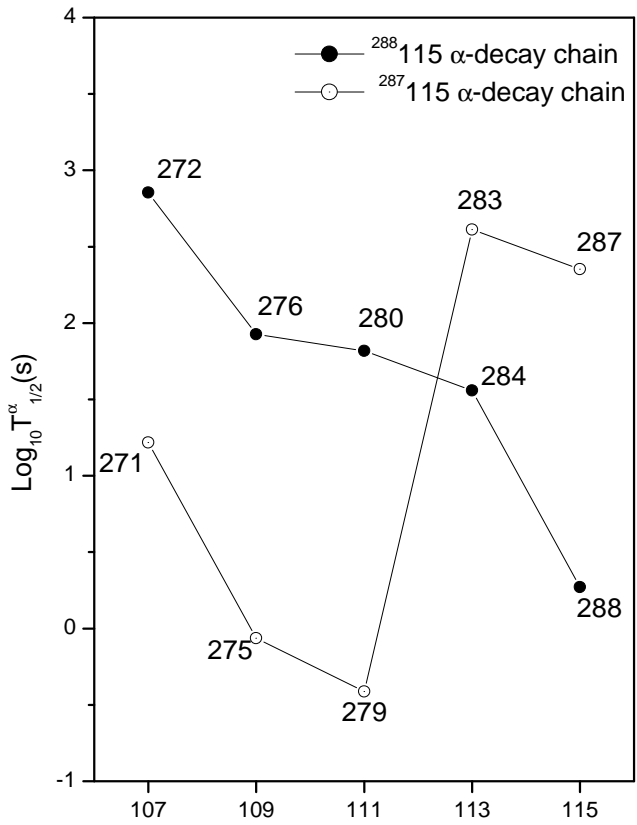
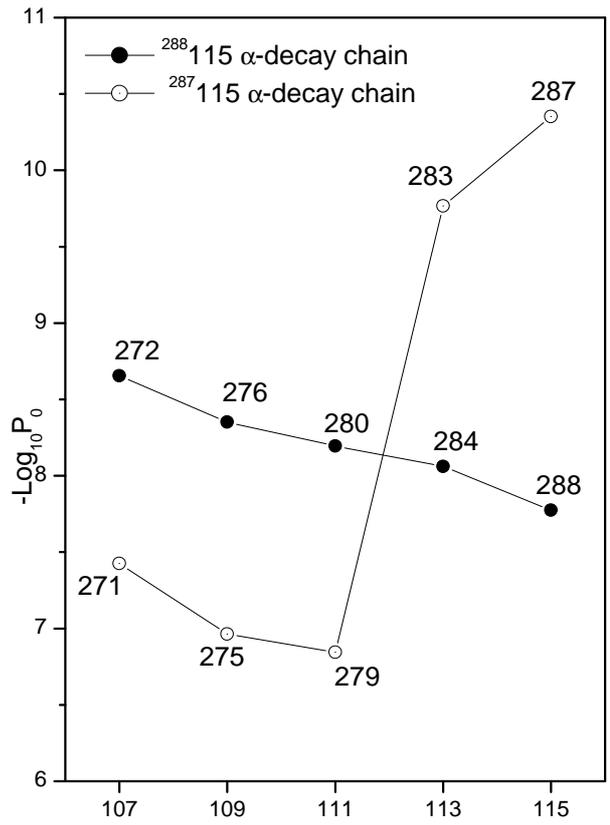
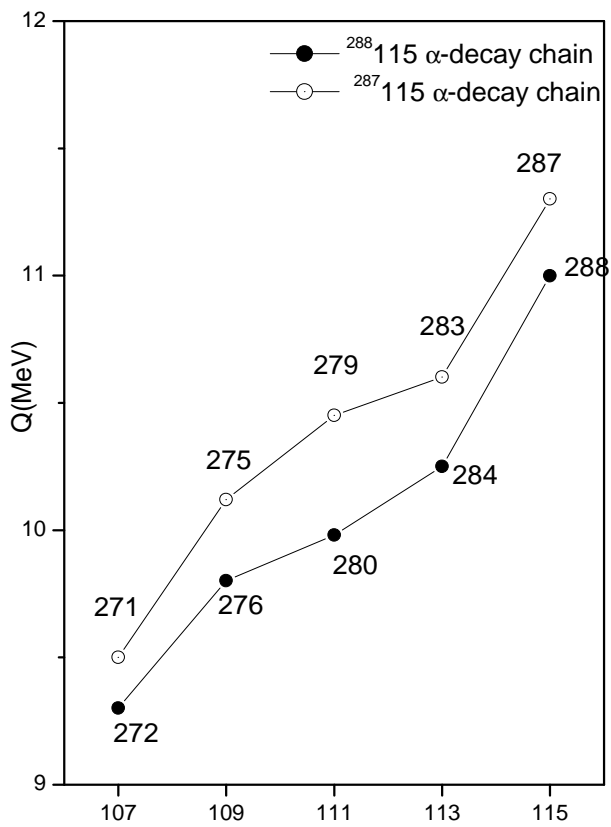
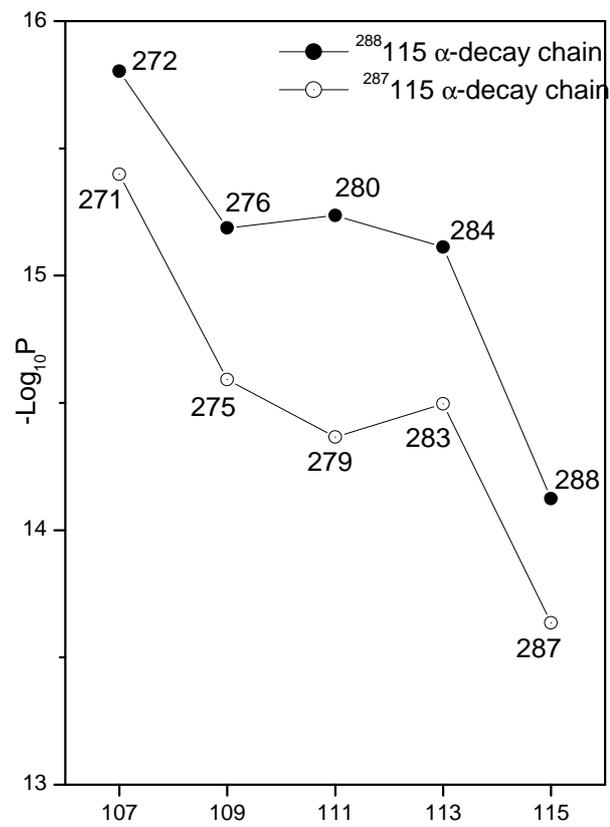
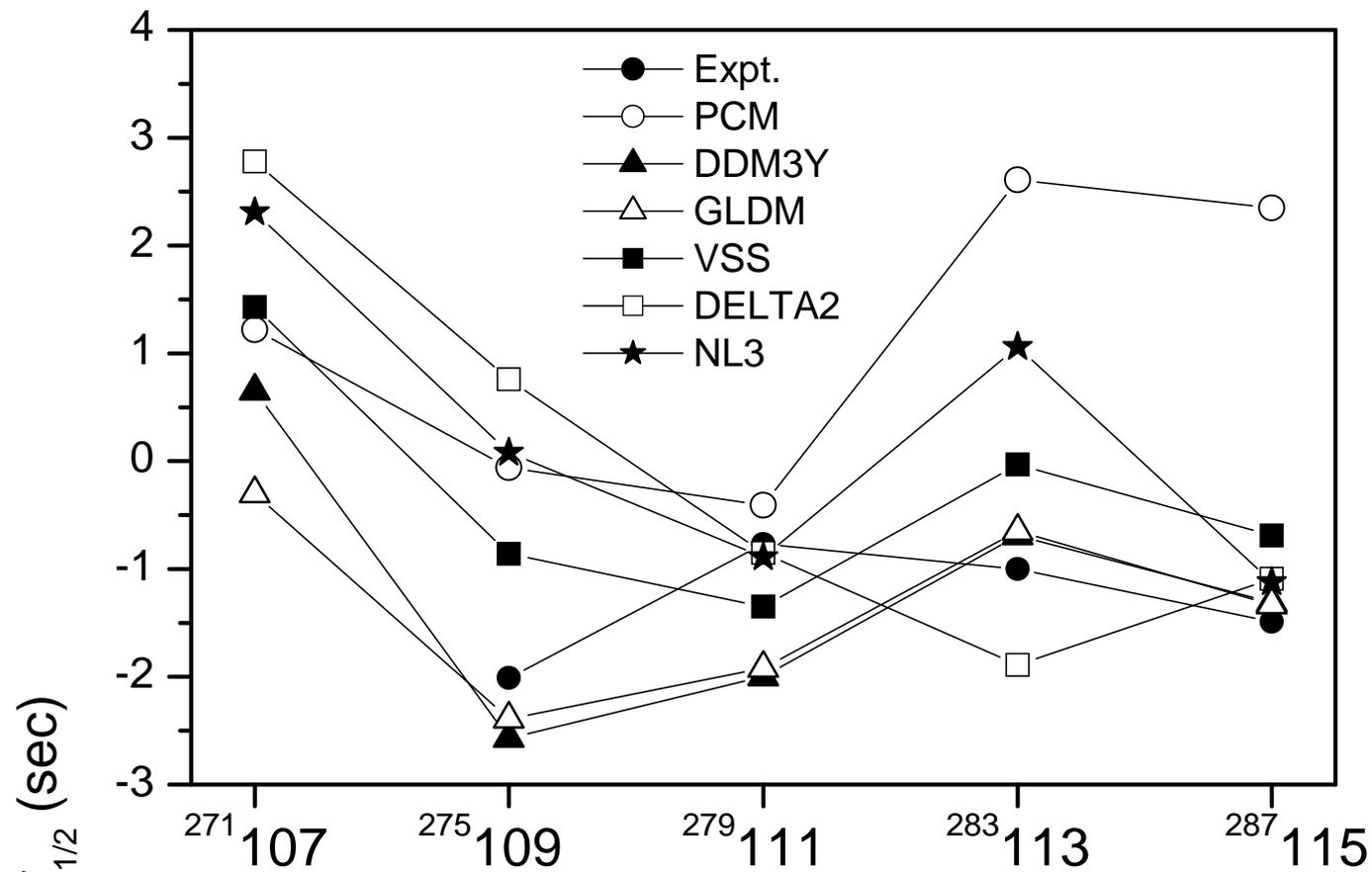
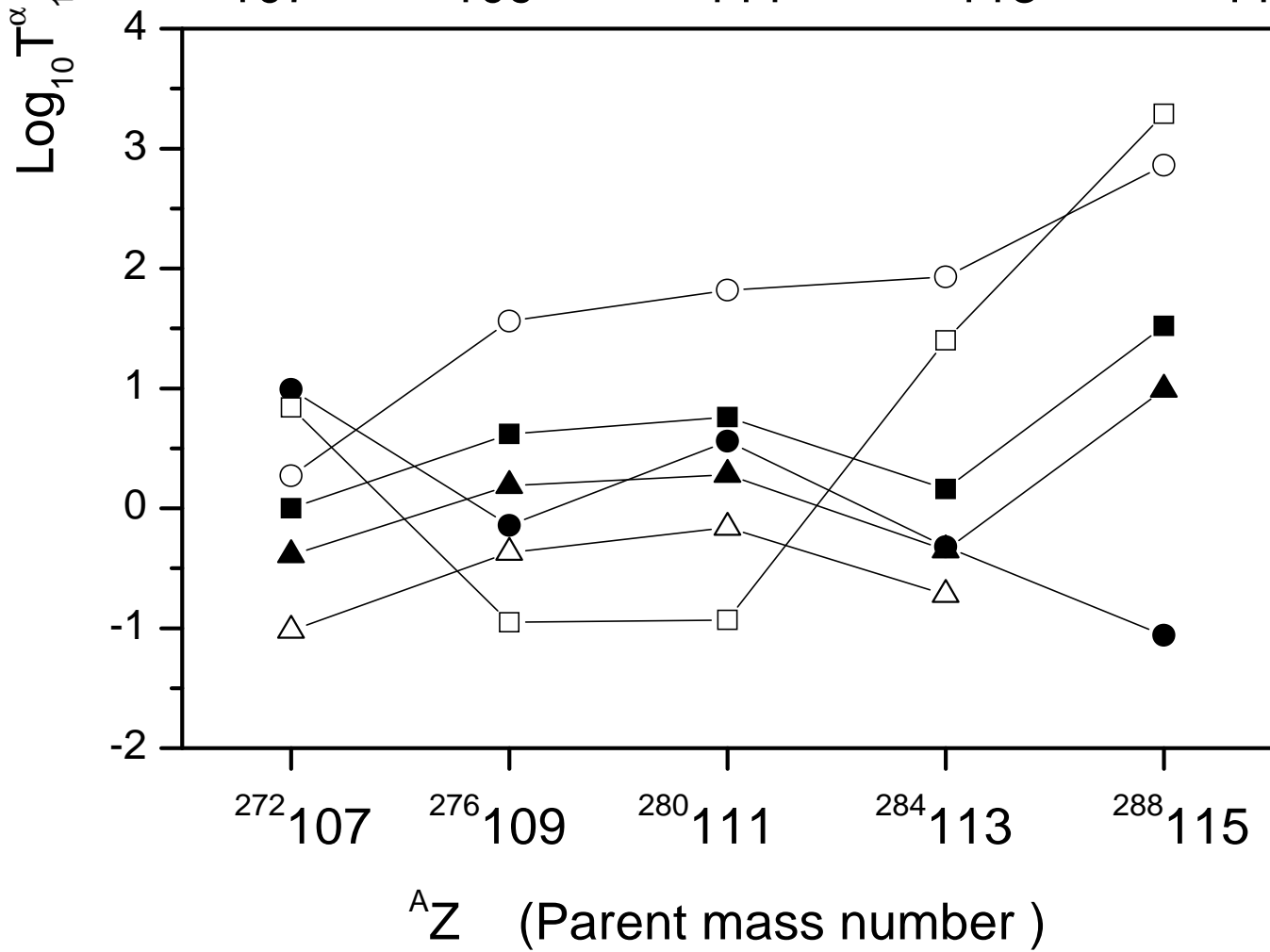
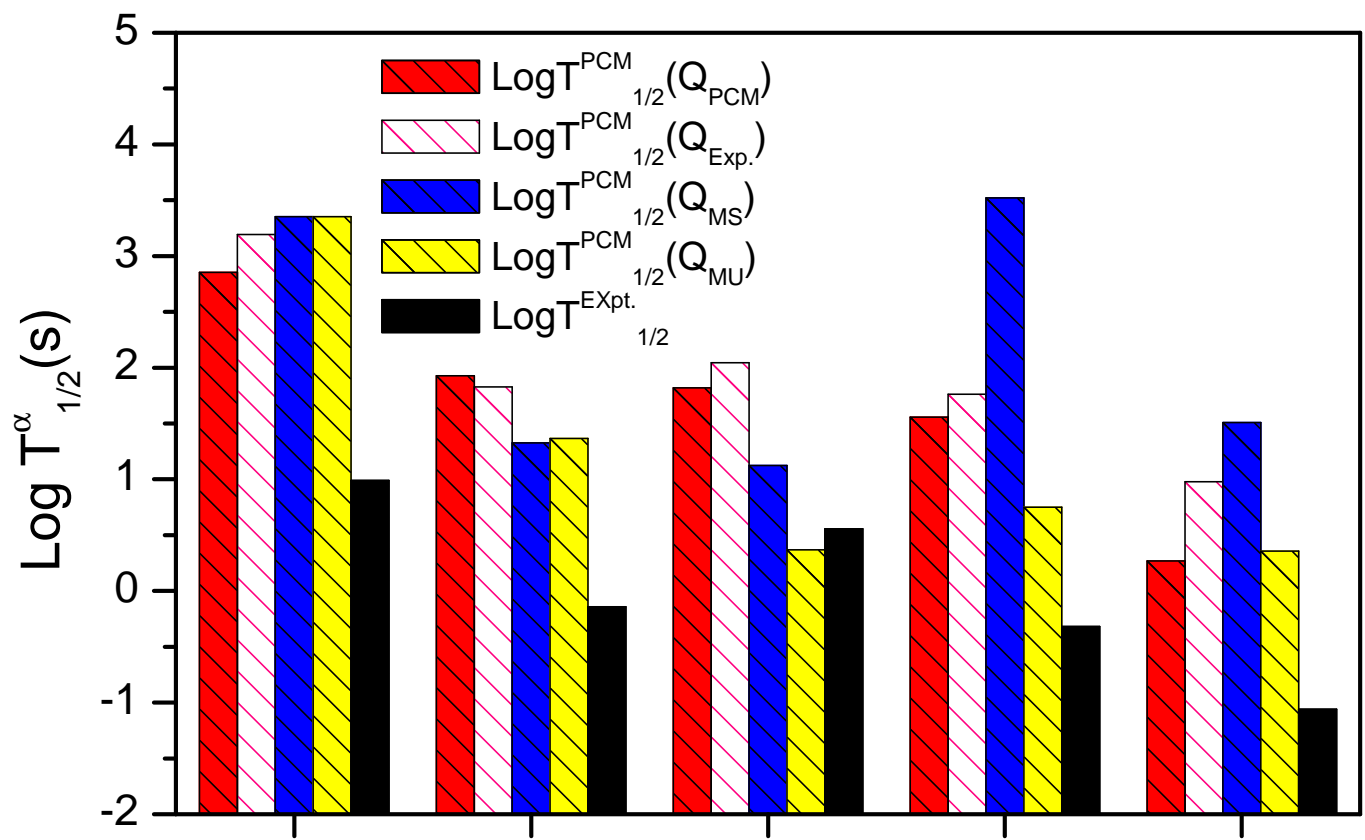
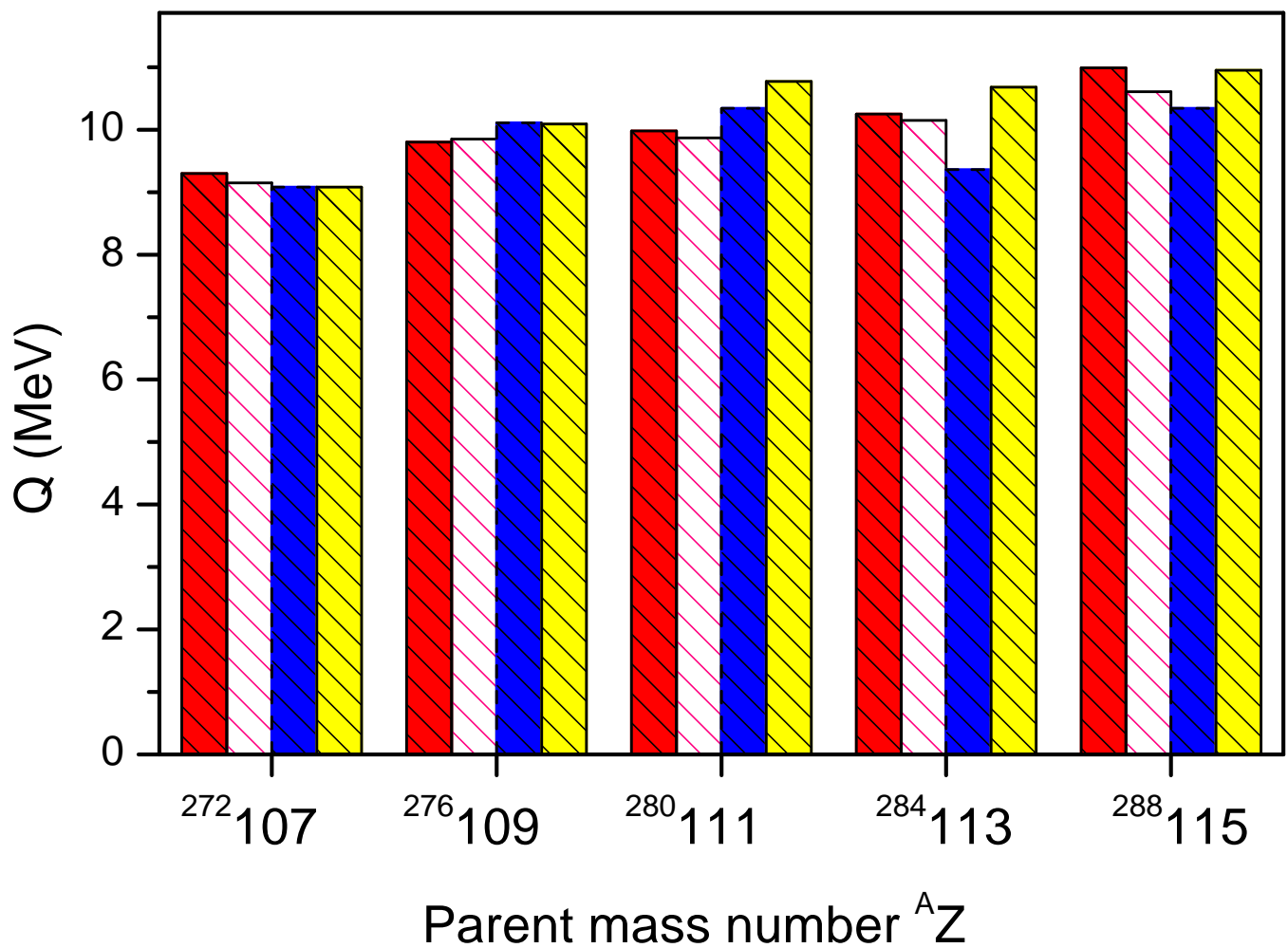
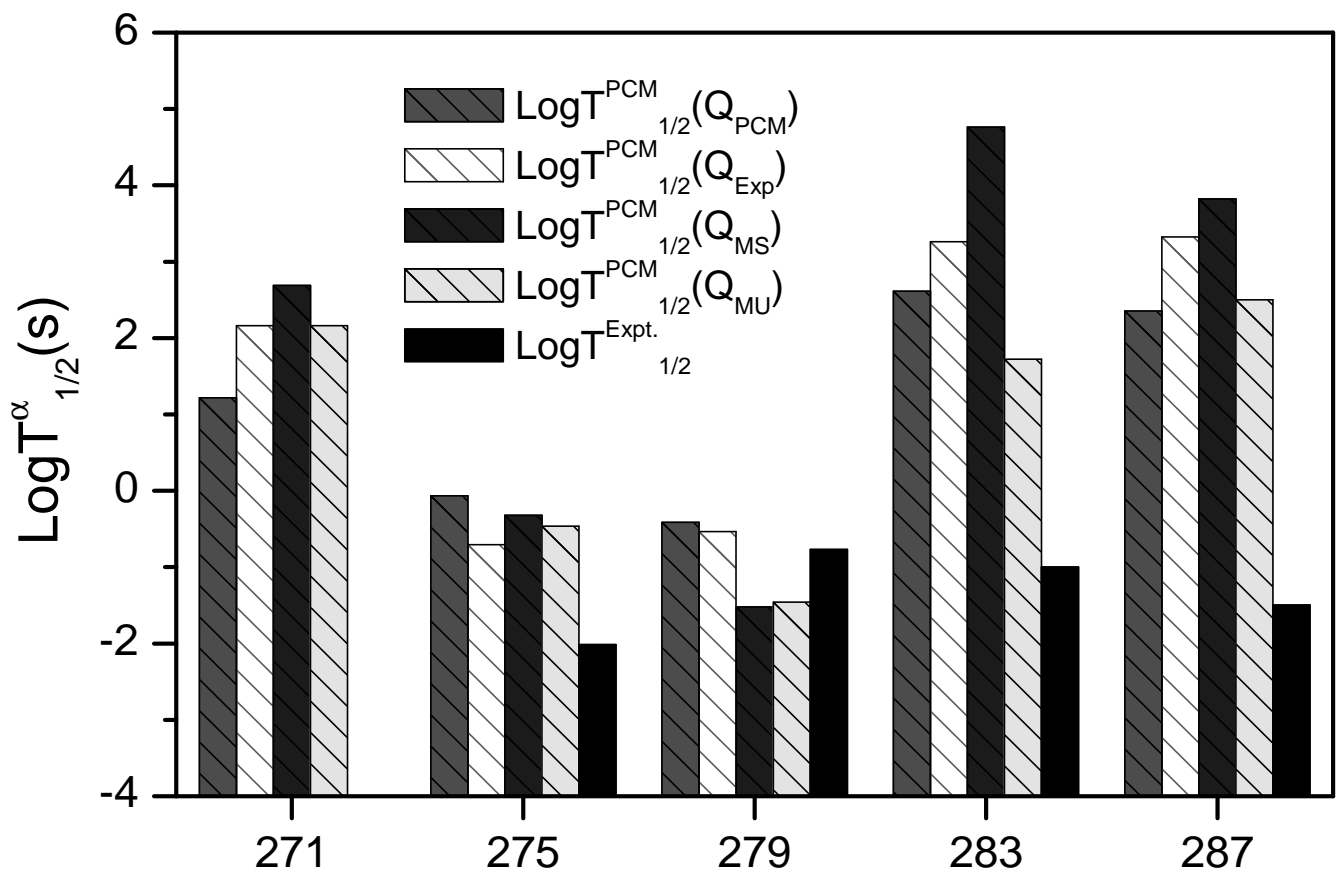
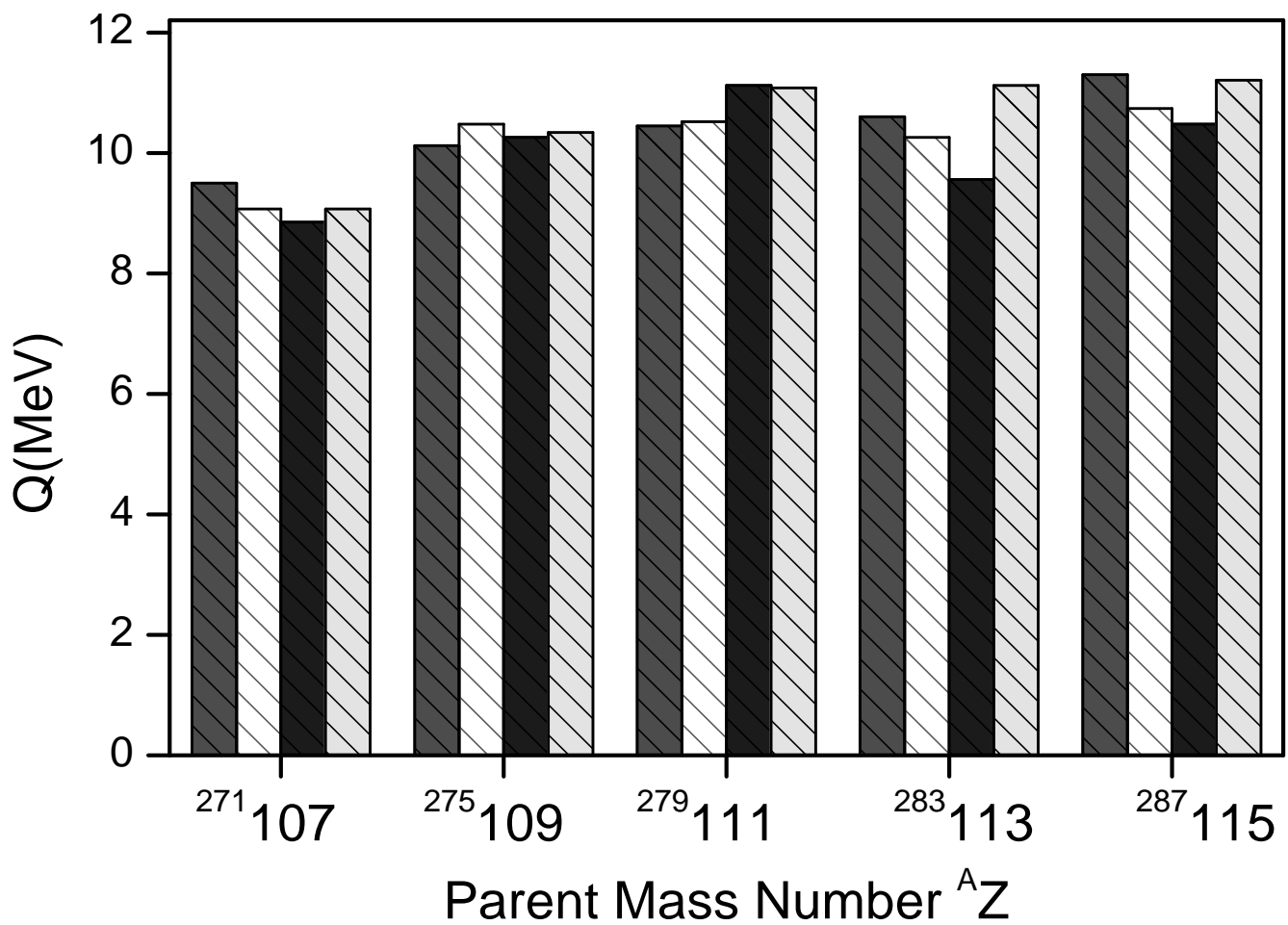
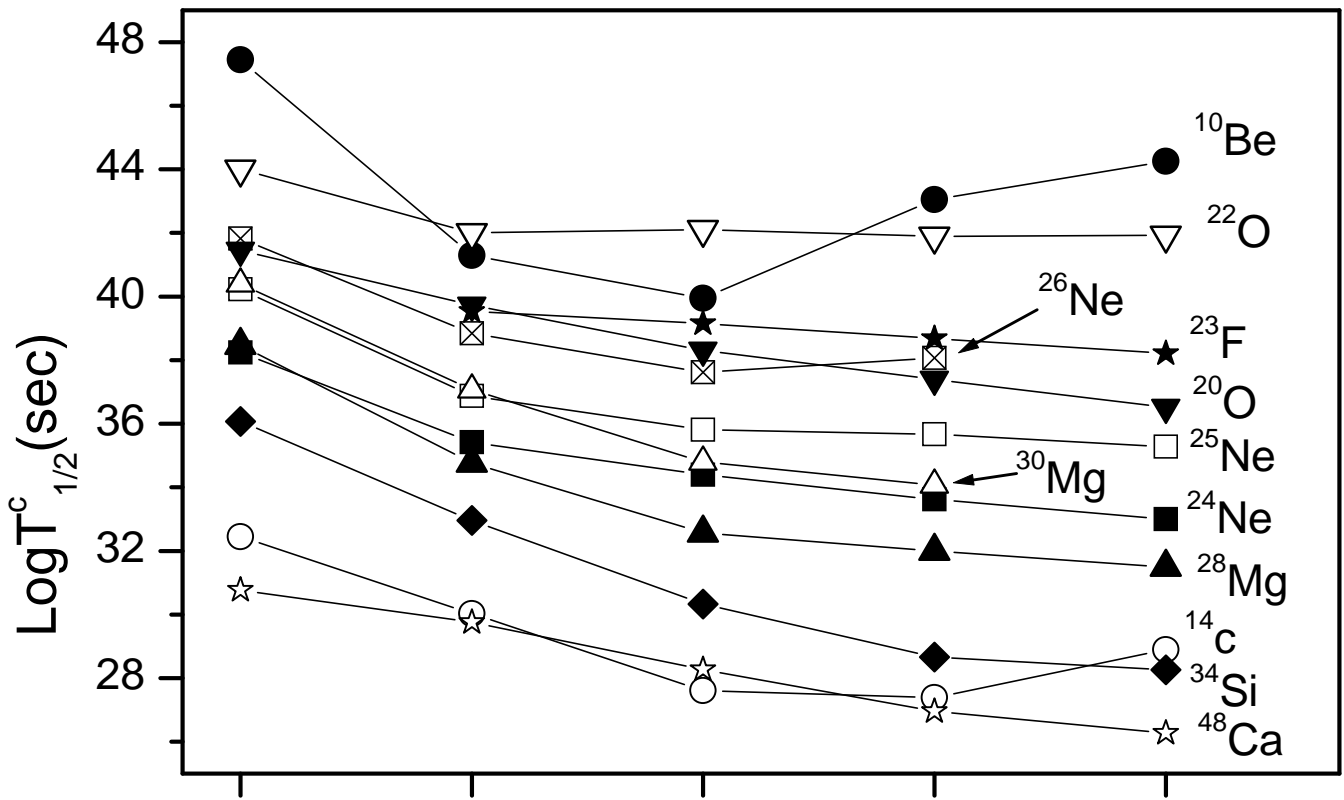
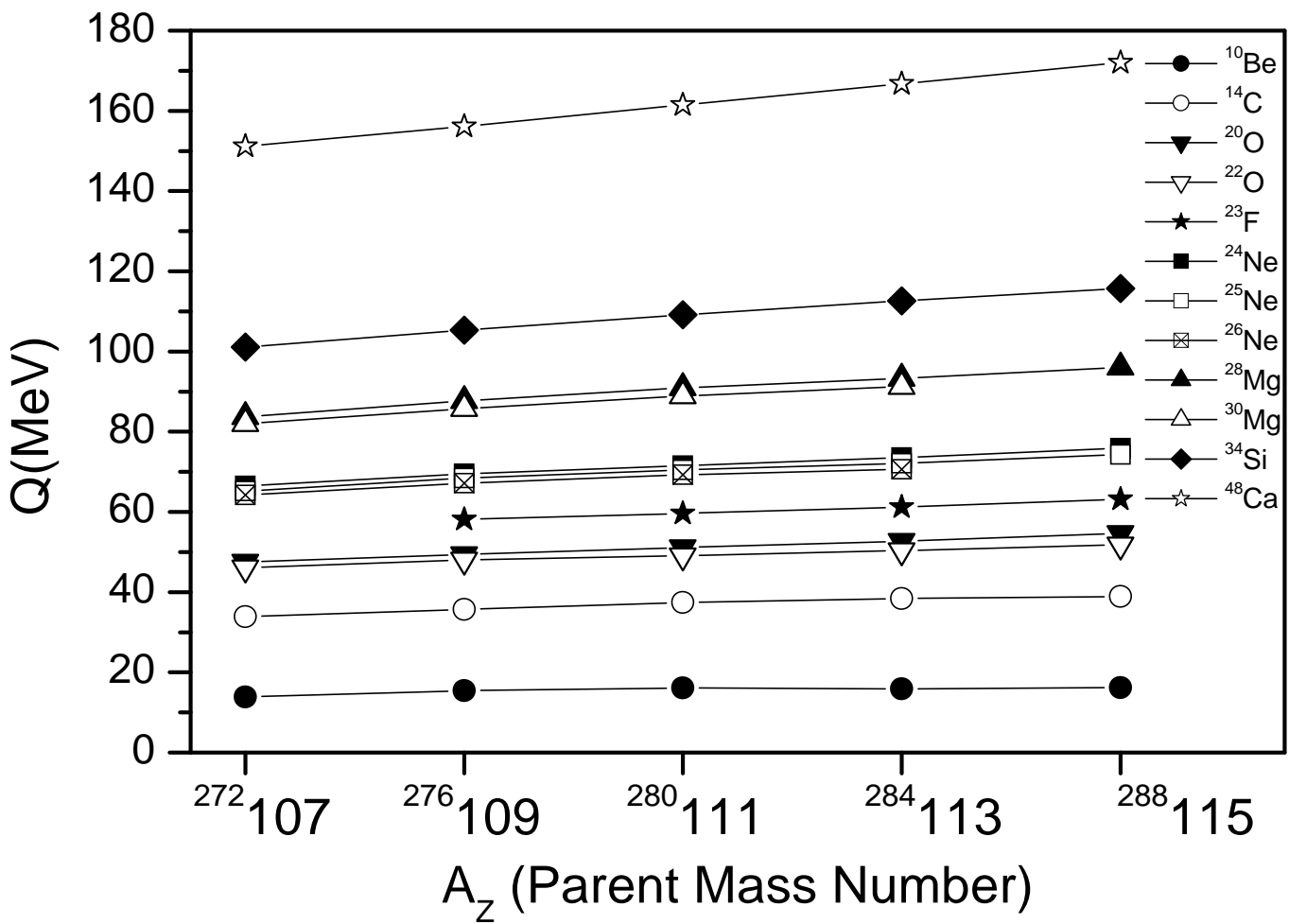
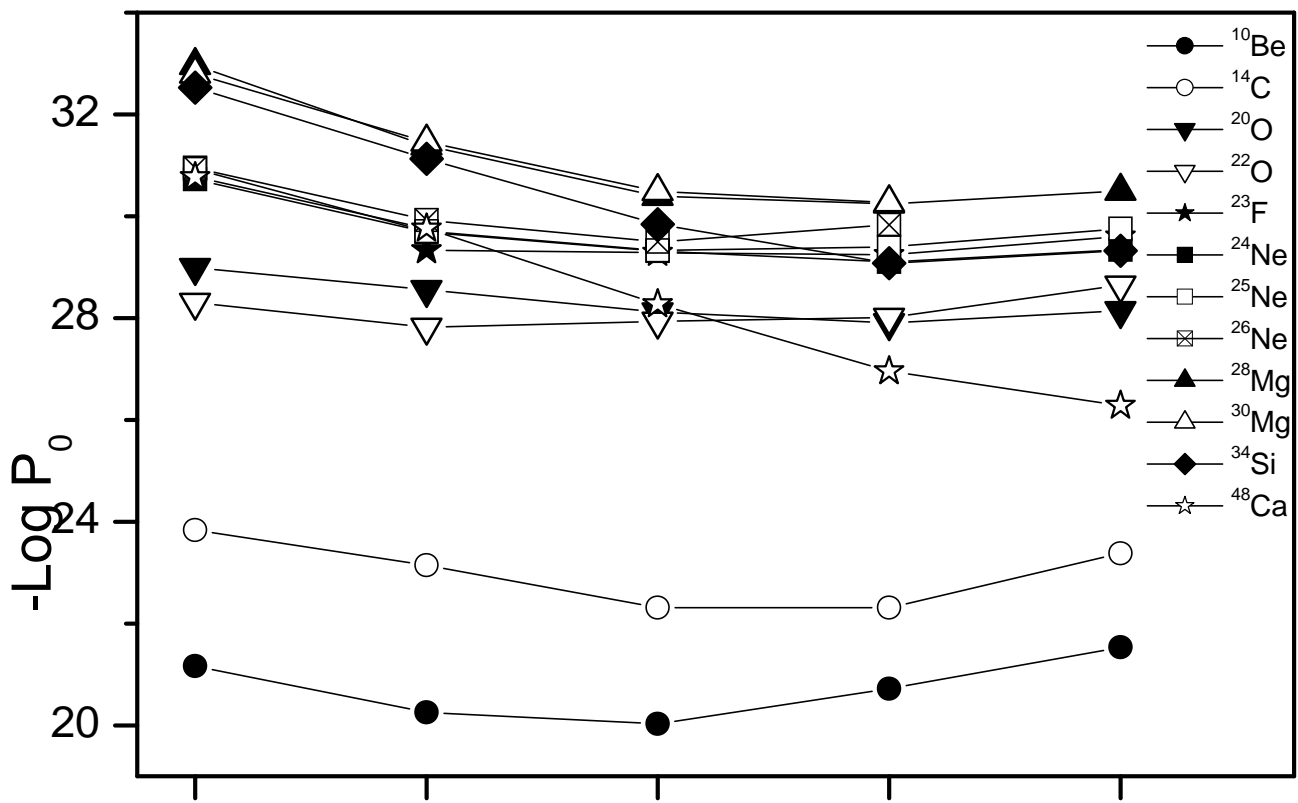
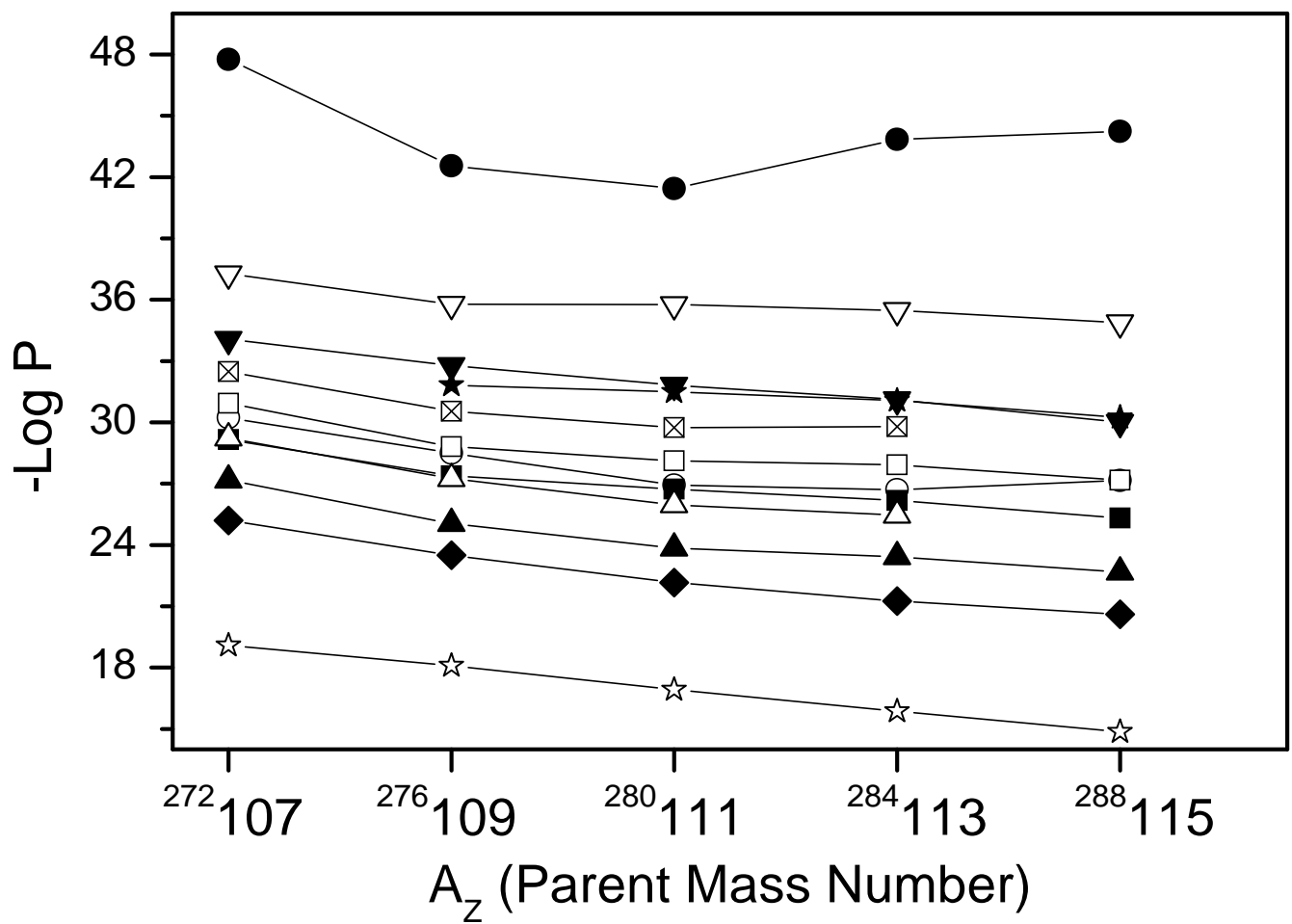
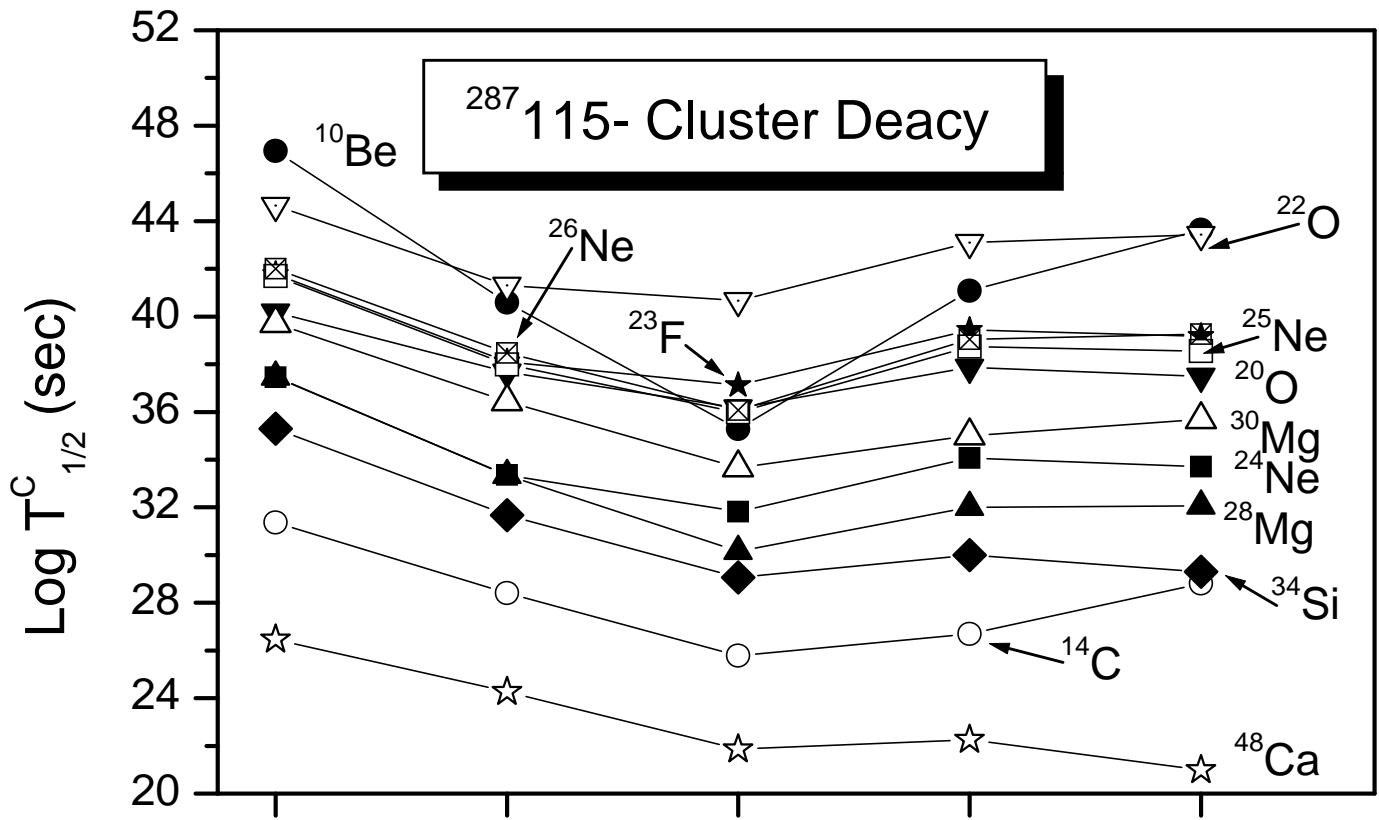
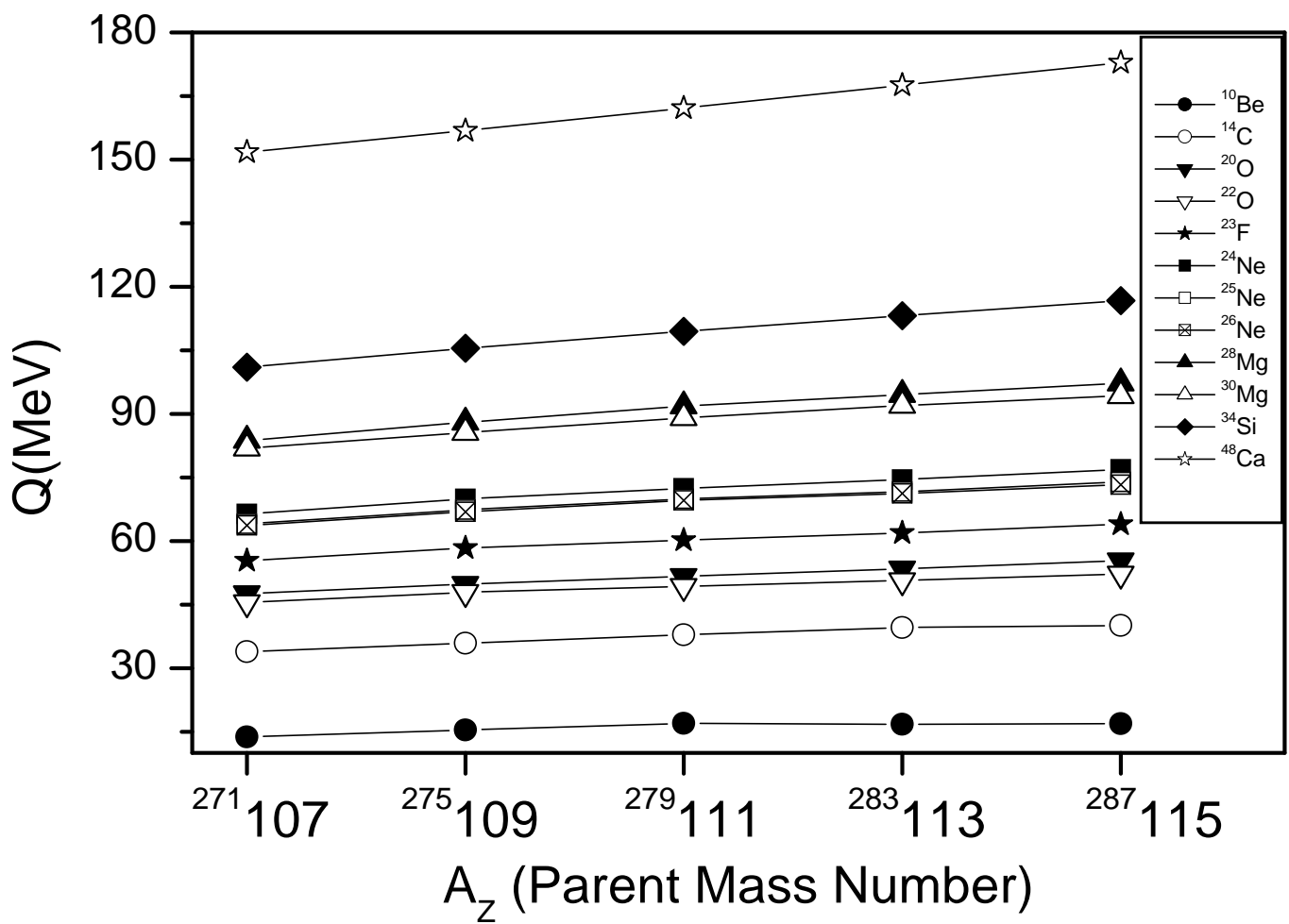
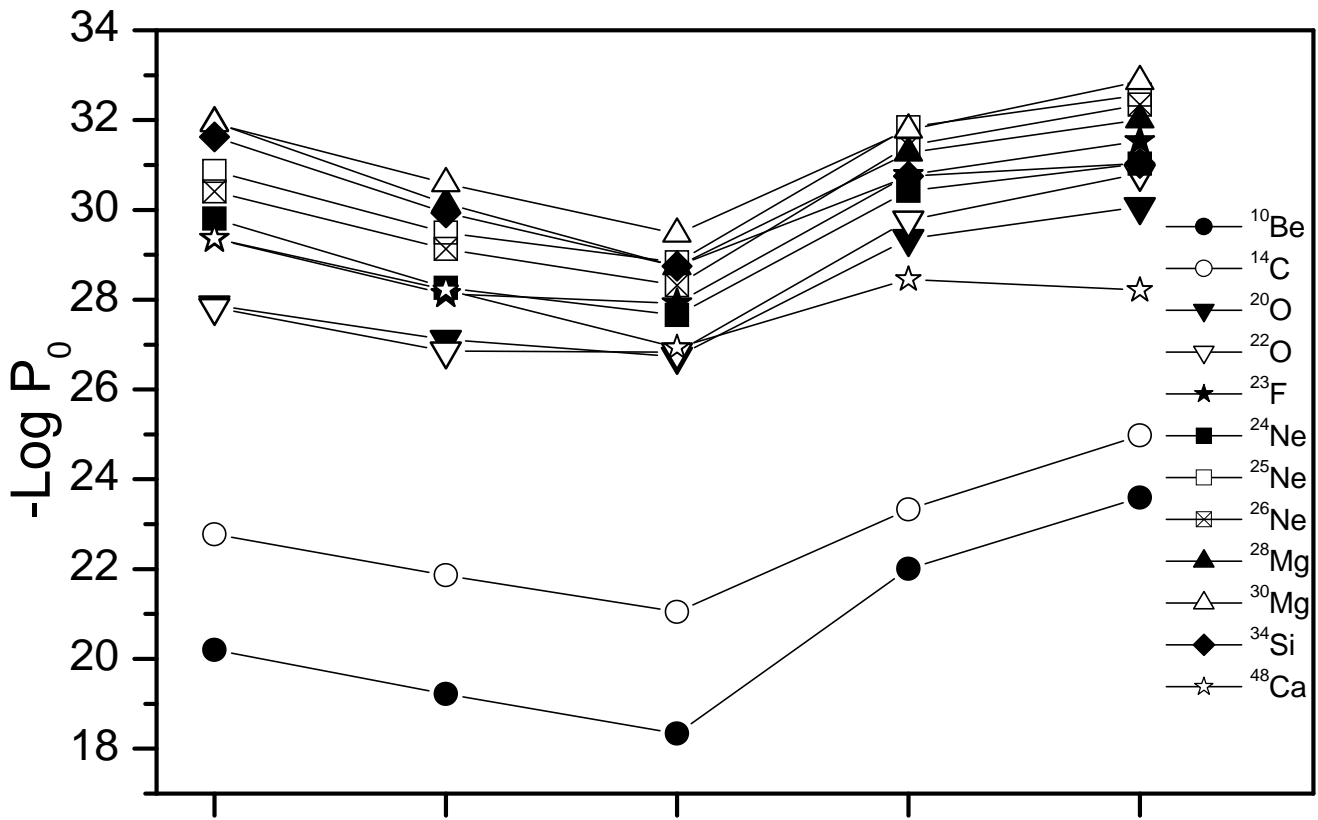
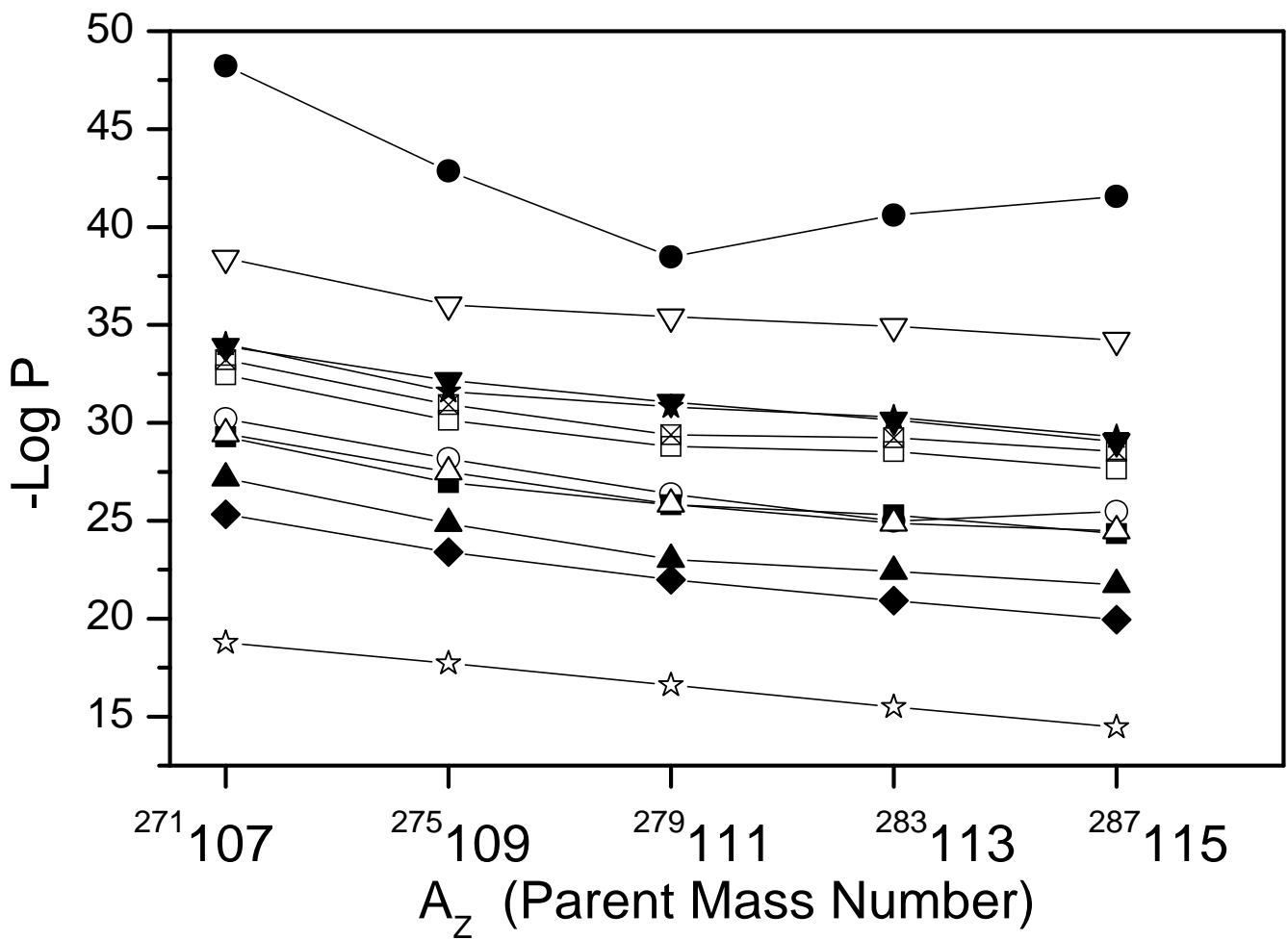
\end{document}